\documentclass[aps,prd,twocolumns,eqsecnum,preprintnumbers,showpacs,amsmath,amssymb]
{revtex4}

\usepackage{graphicx}
\usepackage{dcolumn}
\usepackage{bm}

\begin{document}

\title{NONLINEAR ITERATION SOLUTION FOR THE FULL GLUON PROPAGATOR AS A FUNCTION OF THE MASS GAP}

\author{V. Gogokhia}
\email[]{gogohia@rmki.kfki.hu}

\affiliation{HAS, CRIP, RMKI, Depart. Theor. Phys., Budapest 114,
P.O.B. 49, H-1525, Hungary}

\date{\today}
\begin{abstract}
We have explicitly shown that QCD is the color gauge invariant
theory at non-zero mass gap as well. It has been defined as the
value of the regularized full gluon self-energy at some finite
point. The mass gap is mainly generated by the nonlinear
interaction of massless gluon modes. All this allows one to
establish the structure of the full gluon propagator in the
explicit presence of the mass gap. In this case, the two
independent general types of formal solutions for the full gluon
propagator as a function of the regularized mass gap have been
found. The nonlinear iteration solution at which the gluons remain
massless is explicitly present. The existence of the solution with
an effective gluon mass is also demonstrated.
\end{abstract}

\pacs{ 11.15.Tk, 12.38.Lg}

\keywords{}

\maketitle

\section{Introduction}

Quantum Chromodynamics (QCD) \cite{1,2} is widely accepted as a
realistic quantum field gauge theory of the strong interactions
not only at the fundamental (microscopic) quark-gluon level but at
the hadronic (macroscopic) level as well. This means that in
principle it should describe the properties of experimentally
observed hadrons in terms of experimentally never seen colored
quarks and gluons (the color confinement phenomenon), i.e., to
describe the hadronic world from first principles -- an ultimate
goal of any fundamental theory.

The Lagrangian of QCD, however, does not contain explicitly any of
the mass scale parameters which could have a physical meaning even
after the corresponding renormalization program is performed. This
clearly shows that it is not enough to know it in order to
calculate the physical observables in low-energy QCD from first
principles. It is also important to know the true dynamical
structure of the QCD ground state especially at large distances,
which may be source of the above-mentioned mass scale parameter.
If it will survive the renormalization program, then QCD is a
complete and self-consistent theory without the need to introduce
some extra degrees of freedom in order to generate it. In this way
it may become a mass gap so needed in non-perturbative (NP) QCD in
order to explain the above-mentioned color confinement and other
NP effects \cite{3}. It will be responsible for the NP QCD
dynamics as $\Lambda_{QCD}$ is responsible for the nontrivial
perturbative QCD dynamics (scale violation, asymptotic freedom
(AF) \cite{1,2}).

The propagation of gluons is one of the main dynamical effects in
the QCD vacuum. In our previous work \cite{4} it has been
shown that the only place when the mass gap may appear is the
corresponding Schwinger-Dyson (SD) equation of motion for the full
gluon propagator. It should be complemented by the corresponding
Slavnov-Taylor (ST) identity (see next section). The importance of
this equation is due to the fact that its solutions reflect the
quantum-dynamical structure of the QCD ground state. It is
highly nonlinear (NL) equation, and therefore the number of
independent solutions, which should be considered on equal
footing, is not fixed $a \ priori$. The color gauge structure of
this equation has been investigated in detail in the
above-mentioned paper \cite{4}. We have explicitly shown that the
color gauge invariance of QCD is consistent with the mass gap,
generated in the gluon sector of QCD.

Our primary goal in this investigation is to find formal solutions
for the full gluon propagator as a function of the regularized
mass gap. However, for the sake of completeness and further
clarity, it is instructive to describe briefly the main results of
Ref. \cite{4} in the subsequent section.

\section{The color gauge invariance of QCD at non-zero mass gap}

QCD is a $SU(3)$ color gauge invariant theory. As underlined
above, its dynamical context is determined by the corresponding
equations of motion, among which the SD equation for the full
gluon propagator plays an important role. It can be written as
follows:

\begin{equation}
D_{\mu\nu}(q) = D^0_{\mu\nu}(q) + D^0_{\mu\rho}(q) i
\Pi_{\rho\sigma}(q; D) D_{\sigma\nu}(q),
\end{equation}
where

\begin{equation}
D^0_{\mu\nu}(q) = i \left\{ T_{\mu\nu}(q) + \xi L_{\mu\nu}(q)
\right\} {1 \over q^2}
\end{equation}
is the free gluon propagator, and $\xi$ is the gauge-fixing
parameter. Also, here and everywhere below $T_{\mu\nu}(q) =
\delta_{\mu\nu} - (q_{\mu} q_{\nu} / q^2) = \delta_{\mu\nu} -
L_{\mu\nu}(q)$, as usual. $\Pi_{\rho\sigma}(q; D)$ is the full
gluon self-energy which depends on the full gluon propagator due
to the non-abelian character of QCD. Thus the gluon SD equation is
highly NL one. Evidently, we omit the color group indices, since
for the gluon propagator (and hence for its self-energy) they
factorize, for example $D^{ab}_{\mu\nu}(q) =
D_{\mu\nu}(q)\delta^{ab}$. Diagrammatic representation of the
gluon SD equation (2.1) is shown in our previous work \cite{4}, as
well as the detail description of the full gluon self-energy
$\Pi_{\rho\sigma}(q; D)$. It is the sum of a few terms which are
tensors, having the dimensions of mass squared. All these skeleton
loop integrals are therefore quadratically divergent in
perturbation theory (PT), and so they are assumed to be
regularized, as discussed below. Let us note in advance that here
and below the signature is Euclidean, since it implies $q_i
\rightarrow 0$ when $q^2 \rightarrow 0$ and vice-versa.

\subsection{The mass gap}

Let us introduce the general mass scale parameter $\Delta^2 (D)$,
having the dimensions of mass squared, by the subtraction from the
full gluon self-energy its value at $q=0$. Thus, one obtains

\begin{equation}
\Pi^s_{\rho\sigma}(q; D) = \Pi_{\rho\sigma}(q; D) -
\Pi_{\rho\sigma}(0; D) = \Pi_{\rho\sigma}(q; D) -
\delta_{\rho\sigma} \Delta^2 (D),
\end{equation}
which is nothing but the definition of the subtracted full gluon
self-energy $\Pi^s_{\rho\sigma}(q; D)$. Contrary to QED, QCD being
a non-abelian gauge theory can suffer from infrared (IR)
singularities in the $q^2 \rightarrow 0$ limit due to the
self-interaction of massless gluon modes. Thus the initial
subtraction at zero in the definition (2.3) may be dangerous
\cite{1}. That is why in all the quantities below the dependence
on the finite (slightly different from zero) dimensionless
subtraction point $\alpha$ is to be understood. From a technical
point of view, however, it is convenient to put formally
$\alpha=0$ in all the derivations below, and to restore the
explicit dependence on non-zero $\alpha$ in all the quantities
only at the final stage. At the same time, in all the quantities
where the dependence on $\lambda$ (which is the dimensionless
ultraviolet (UV) regulating parameter) and $\alpha$ is not shown
explicitly, nevertheless, it should be assumed. For example,
$\Delta^2(D) \equiv \Delta^2(\lambda, \alpha; D)$ and similarly
for all other quantities. So all the expressions are regularized.
For our purpose, in principle, it is not important how $\lambda$
and $\alpha$ have been introduced. They should be removed at the
final stage only as a result of the renormalization program.

By the mass gap we understand some fixed mass squared which is
related to $\Delta^2 (D)$ as follows:

\begin{equation}
\Delta^2 (D) = \Delta^2 \times c(D),
\end{equation}
where the dimensionless constant $c(D)$ depends on $D$, while the
fixed mass squared $\Delta^2$ does not depend on $D$. It will be
called the mass gap. As the general mass scale parameter itself
and constant $c(D)$, it may depend on all other dimensionless
parameters of the theory, namely $\Delta^2 \equiv
\Delta^2(\lambda, \alpha, \xi, g^2)$, where $g^2$ is the coupling
constant squared and so on. In this section we will not
distinguish between $\Delta^2(D)$ and $\Delta^2$, calling both the
mass gap, for simplicity. From the subtraction (2.3) it follows
that the mass gap $\Delta^2$, having the dimensions of mass
squared, is dynamically generated in the QCD gluon sector. It is
defined as the value of the full gluon self-energy at some finite
point (see discussion above). It is mainly due to the nonlinear
interaction of massless gluon modes. Let us remind that no
truncations/approximations/assumptions/, as well as no special
gauge choice are made for the regularized skeleton loop integrals,
contributing to the full gluon self-energy \cite{4}.

\subsection{The transversality of the full gluon self-energy}

Let us continue with the general decompositions of the full gluon
self-energy and its subtracted counterpart, which enter the
subtraction (2.3), as follows:

\begin{eqnarray}
\Pi_{\rho\sigma}(q; D) &=&  T_{\rho\sigma}(q) q^2 \Pi(q^2; D) +
q_{\rho} q_{\sigma} \tilde{\Pi}(q^2; D), \nonumber\\
\Pi^s_{\rho\sigma}(q; D) &=&  T_{\rho\sigma}(q) q^2 \Pi^s(q^2; D)
+ q_{\rho} q_{\sigma} \tilde{\Pi}^s(q^2; D),
\end{eqnarray}
where all the invariant functions of $q^2$ are dimensionless ones,
while in addition the invariant functions $\Pi^s(q^2; D)$ and
$\tilde{\Pi}^s(q^2; D)$ cannot have the pole-type singularities in
the $q^2 \rightarrow 0$ limit, since $\Pi^s_{\rho\sigma}(0; D)
=0$, by definition; otherwise they remain arbitrary.

Contracting them with $q_{\rho}$ along with the subtraction (2.3),
one obtains

\begin{equation}
\tilde{\Pi}(q^2; D) = \tilde{\Pi}^s(q^2; D) + {\Delta^2(D) \over
q^2},
\end{equation}
and

\begin{equation}
\Pi(q^2; D) = \Pi^s(q^2; D) + {\Delta^2(D) \over q^2}.
\end{equation}

It is worth emphasizing that the full gluon self-energy has a
massless single particle singularity due to non-zero mass gap
$\Delta^2(D)$, which is of the non-perturbative (NP) origin. At
the same time, its subtracted counterpart cannot have such a
singularity, as mentioned above. In other words, this means that
in the explicit presence of the mass gap both invariant functions
of the full gluon self-energy gain additional contributions due to
it (of course, not only at some finite subtraction point $q^2 =
\mu^2 \neq 0$). If the mass gap is welcome in the transversal
invariant function $\Pi(q^2; D)$, it is not welcome in its
longitudinal counterpart $\tilde{\Pi}(q^2; D)$, since just it
violates the ST identity. Let us also note in advance that
transversality of the full gluon self-energy and its subtracted
counterpart can be achieved only in the formal $\Delta^2(D)=0$
limit (for a brief discussion of all these preliminary remarks see
subsections below). So in the general case of non-zero
$\Delta^2(D)$ only two possibilities remain.

(i). Both are not transversal and then

\begin{eqnarray}
q_{\rho} \Pi_{\rho\sigma}(q; D) &=& q_{\sigma} q^2
\tilde{\Pi}(q^2; D) = q_{\sigma} [q^2 \tilde{\Pi}^s(q^2; D)
+ \Delta^2(D)] \neq 0, \nonumber\\
q_{\rho} \Pi^s_{\rho\sigma}(q; D) &=& q_{\sigma} q^2
\tilde{\Pi}^s(q^2; D) = q_{\sigma} [q^2 \tilde{\Pi}(q^2; D) -
\Delta^2(D)].
\end{eqnarray}
The last inequality in the first of the relations (2.8) follows
from the fact that $\tilde{\Pi}^s(q^2; D)$ cannot have a single
particle singularity $ - \Delta^2(D)/ q^2$ in order to cancel
$\Delta^2(D)$.

(ii). Transversality of the subtracted gluon self-energy is
maintained, i.e., $\tilde{\Pi}^s(q^2; D)=0$ and then

\begin{equation}
q_{\rho} \Pi_{\rho\sigma}(q; D) = q_{\sigma} q^2 \tilde{\Pi}(q^2;
D) = q_{\sigma} \Delta^2(D) \neq 0, \quad  q_{\rho}
\Pi^s_{\rho\sigma}(q; D) = 0,
\end{equation}
Contrary to the first case, now we know how precisely the
transversality of the full gluon self-energy is violated. So it is
always violated at non-zero mass scale parameter $\Delta^2(D)$. In
this connection one thing should be made perfectly clear. It is
the initial subtraction (2.3) which leaves the subtracted
gluon-self energy logarithmical divergent only, and hence the
invariant function $\Pi^s(q^2; D)$ is free of the quadratic
divergences, but a logarithmic ones can be still present in it, at
any $D$. Since the transversality condition for the full gluon
self-energy is violated in these relations, that is why we cannot
disregard $\Delta^2(D)$ from the very beginning (compare with the
pure quark case considered in our initial work \cite{4}).

\subsection{The ST identity for the full gluon propagator}

In order to calculate the physical observables in QCD from first
principles, we need the full gluon propagator rather than the full
gluon self-energy. The basic relation to which the full gluon
propagator should satisfy is the corresponding ST identity

\begin{equation}
q_{\mu}q_{\nu} D_{\mu\nu}(q) = i \xi.
\end{equation}
It is a consequence of the color gauge invarince/symmetry of QCD,
and therefore "is an exact constraint on any solution to QCD"
\cite{1}. This is true for any other ST identities. Being a result
of this exact symmetry, it is the general one, and it is important
for the renormalization of QCD. If some equation, relation or the
regularization scheme, etc.  do not satisfy it automatically,
i.e., without any additional conditions, then they should be
modified and not this identity (identity is an equality, where
both sides are the same, i.e., there is no room for additional
conditions). In other words, all the relations, equations,
regularization schemes, etc. should be adjusted to it and not vice
versa. It implies that the general tensor decomposition of the
full gluon propagator is

\begin{equation}
D_{\mu\nu}(q) = i \left\{ T_{\mu\nu}(q) d(q^2) + \xi L_{\mu\nu}(q)
\right\} {1 \over q^2},
\end{equation}
where the invariant function $d(q^2)$ is the corresponding Lorentz
structure of the full gluon propagator (sometimes we will call it
as the full effective charge ("running"), for simplicity). Let us
emphasize once more that these basic relations are to be satisfied
in any case, for example, whether the mass gap or any other mass
scale parameter is put formally zero or not.

On account of the exact relations (2.5), (2.6) and (2.7), the
initial gluon SD equation (2.1) can be equivalently re-written
down as follows:

\begin{equation}
D_{\mu\nu}(q) = D^0_{\mu\nu}(q) + D^0_{\mu\rho}(q)i
T_{\rho\sigma}(q) [q^2 \Pi^s(q^2; D) + \Delta^2(D)]
D_{\sigma\nu}(q) + D^0_{\mu\rho}(q)i L_{\rho\sigma}(q) q^2
\tilde{\Pi}(q^2; D) D_{\sigma\nu}(q).
\end{equation}

Contracting this equation with $q_{\mu}$ and $q_{\nu}$, one
arrives at $q_{\mu}q_{\nu} D_{\mu\nu}(q) = i \xi - i \xi^2
\tilde{\Pi}(q^2; D)$, so the ST identity (2.10) is not
automatically satisfied. In order to get from this relation the ST
identity, one needs to put $\tilde{\Pi}(q^2; D) =0$, which is
equivalent to $\tilde{\Pi}^s(q^2; D) = - (\Delta^2(D) / q^2)$, as
it follows from the relation (2.6). This, however, is impossible
since $\tilde{\Pi}^s(q^2; D)$ cannot have the power-type
singularities at small $q^2$, as underlined above. The only
solution to the previous relation is to disregard $\Delta^2(D)$
from the very beginning, i.e., put formally zero $\Delta^2(D) =0$
everywhere. In this case from all the relations it follows that
the gluon full self-energy coincides with its subtracted
counterpart, and both quantities become purely transversal, i.e.,
$\Pi(q^2; D) = \Pi^s(q^2; D)$ and $\tilde{\Pi}(q^2; D) =
\tilde{\Pi}^s(q^2; D)=0$ (see relations (2.5)-(2.7)).

The one way to satisfy the ST identity and thus to maintain the
color gauge structure of QCD is to discard the mass gap
$\Delta^2(D)$ from the very beginning, i.e., put it formally zero
$\Delta^2(D)=0$ in all the equations, relations, etc. In this
limit the initial gluon SD equation (2.12) is modified to

\begin{equation}
D^{PT}_{\mu\nu}(q) = D^0_{\mu\nu}(q) + D^0_{\mu\rho}(q)i
T_{\rho\sigma}(q) q^2 \Pi^s(q^2; D^{PT})D^{PT}_{\sigma\nu}(q),
\end{equation}
and the corresponding Lorentz structure which appears in Eq.
(2.11) becomes

\begin{equation}
d^{PT}(q^2) = { 1 \over 1 + \Pi^s(q^2; D^{PT})}.
\end{equation}
It is easy to see that the gluon SD equation (2.13) automatically
satisfies the ST identity (2.10) now. Evidently, in the formal
$\Delta^2(D)=0$ limit we denote $D_{\mu\nu}(q)$ and $d(q^2)$ as
$D^{PT}_{\mu\nu}(q)$ and $d^{PT}(q^2)$, respectively (for reason
see below). As it has been pointed out in Ref. \cite{4}, in this
case there will be no problems for ghosts to accomplish their
role, namely to cancel the longitudinal component in the full
gluon propagator (2.13).

\subsection{The general structure of the full gluon propagator}

The formal $\Delta^2(D)=0$ limit is a real way how to preserve the
color gauge invariance in QCD. Then a natural question arises why
does the mass gap $\Delta^2(D)$ exist in this theory at all? There
is no doubt that the color gauge invariance of QCD should be
maintained at non-zero mass gap as well, since it is explicitly
present in the full gluon self-energy, and hence in the full gluon
propagator. However, by keeping it "alive", the two important
problems arise. The first problem is how to replace the original
gluon SD equation (2.12), since it is not consistent with the ST
identity unless the mass gap is discarded from the very beginning
(see above). The second problem is how to make the full gluon
propagator purely transversal when the mass gap is explicitly
present.

By introducing the spurious technics we were able to show that the
ST identity (2.10) can be automatically satisfied at non-zero mass
gap $\Delta^2(D)$ as well. In other words, our aim is to save the
mass gap in the transversal invariant function (2.7), while
removing it from the longitudinal invariant function (2.6), but
without going formally to the PT $\Delta^2=0$ limit. In order to
keep the mass gap "alive", and, at the same time, to satisfy the
ST identity (2.10), we introduced a temporary dependence on
$\Delta^2(D)$ in the free gluon propagator, thus making it an
auxiliary (spurious) free gluon propagator. Substituting it into
the initial gluon SD equation (2.12) and restoring again the
dependence on the free gluon propagator, such obtained gluon SD
equation should satisfy the ST identity (2.10). After doing some
tedious algebra, one finally obtains \cite{4}

\begin{equation}
D_{\mu\nu}(q) = D^0_{\mu\nu}(q) + D^0_{\mu\rho}(q)i
T_{\rho\sigma}(q) [q^2 \Pi^s(q^2; D) + \Delta^2(D)]
D_{\sigma\nu}(q).
\end{equation}
Such modified gluon SD equation (2.15) is satisfied by the same
expression for the Lorentz structure $d(q^2)$ in Eq.~(2.11) as the
original gluon SD equation (2.12), namely

\begin{equation}
d(q^2) = {1 \over 1 + \Pi^s(q^2; D) + (\Delta^2(D) / q^2)},
\end{equation}
which is not surprising, since the original gluon SD equation
(2.12) and its modified version (2.15) differ from each other only
by the longitudinal (unphysical) part.

{\bf However, the important observation is that now it is not
required to put the mass gap $\Delta^2(D)$ formally zero
everywhere}. The spurious mechanism does not affect the dynamical
context of the original gluon SD equation. In other words, it
makes it possible to retain the mass gap in the transversal part
of the gluon SD equation, and, at the same time, to cancel the
term in its longitudinal part, which violates the ST identity. In
this way, the modified gluon SD equation (2.15) satisfies
automatically the ST identity (2.10).

Due to AF in QCD the PT regime is realized at $q^2 \rightarrow
\infty$. In this limit all the Green's functions are possible to
approximate by their free PT counterparts (up to the corresponding
PT logarithms). However, from the relation (2.16) it follows that
in this limit the mass gap term contribution $\Delta^2(D) / q^2$
is only next-to-next-to-leading order one. The leading order
contribution is the subtracted gluon self-energy $\Pi^s(q^2; D)$,
which behaves like $\ln q^2$ in this limit, as mentioned above.
The constant $1$ is the next-to-leading order term in the $q^2
\rightarrow \infty$ limit. Such a special structure of the
relation (2.16), namely the mass gap enters it through the
combination $\Delta^2(D) / q^2$ in its denominator only, explains
immediately why the mass gap $\Delta^2(D)$ is not important in PT.
From this structure it follows that the PT regime at $q^2
\rightarrow \infty$ is effectively equivalent to the formal $\Delta^2(D)=0$
limit and vice versa. That is the reason why this limit can be
called the PT limit. And that is why we denote $D_{\mu\nu}(q;
\Delta^2=0) = D_{\mu\nu}(q; 0) \equiv D^{PT}_{\mu\nu}(q)$, and
hence $d(q^2; \Delta^2=0) = d(q^2; 0) \equiv d^{PT}(q^2)$, etc.,
in accordance with the previous notations.
Let us note, however, that sometimes it is useful to distinguish
between the asymptotic suppression of the mass gap contribution $\Delta^2 / q^2$ in the
$q^2 \rightarrow \infty$ limit and the formal PT $\Delta^2=0$ limit (see our subsequent paper).

Thus the formal PT $\Delta^2(D)=0$ limit exists, and it is a
regular one. As it follows from above, in this limit one recovers
the PT QCD system of equations (2.13)-(2.14) from the NP QCD one
(2.15)-(2.16). So, we distinguish between the PT and NP phases in
QCD by the explicit presence of the mass gap. Its aim is to be
responsible for the NP QCD dynamics, since it dominates at $q^2
\rightarrow 0$ in the "solution" (2.16). When it is put formally
zero, then the PT phase survives only. Evidently, when such a
scale is explicitly present then the QCD coupling constant plays
no role in the NP QCD dynamics.

\subsection{Transversality of the relevant full gluon propagator}

The NP QCD system of equations(2.15)-(2.16) depends explicitly on
the mass gap $\Delta^2$. As it has been discussed in detail in our
previous work \cite{4}, then the ghosts are not able to cancel the
longitudinal component in the full gluon propagator, i.e., they
are of no use in this case (the transversality condition for the
full gluon self-energy is always violated, see relations (2.8) and
(2.9)). This is the price we have paid to keep the mass gap
"alive" in the full gluon propagator. Our aim here is to formulate
a method which allows one to make the gluon propagator, relevant
for NP QCD, purely transversal in a gauge invariant way, even if
the mass gap is explicitly present.

For this purpose let us define the truly NP (TNP) part of the full
gluon propagator as follows:

\begin{equation}
D^{TNP}_{\mu\nu}(q; \Delta^2) = D_{\mu\nu}(q; \Delta^2) -
D_{\mu\nu}(q; \Delta^2=0) = D_{\mu\nu}(q; \Delta^2) -
D^{PT}_{\mu\nu}(q),
\end{equation}
i.e., the subtraction is made with respect to the mass gap
$\Delta^2$, and therefore the separation between these two terms
is exact. So it becomes

\begin{equation}
D^{TNP}_{\mu\nu}(q; \Delta^2) = i T_{\mu\nu} (q) \Bigr[ d(q^2;
\Delta^2) -  d^{PT}(q^2) \Bigl] {1 \over q^2} = i T_{\mu\nu} (q)
d^{TNP}(q^2; \Delta^2) {1 \over q^2},
\end{equation}
where the explicit expression for the TNP Lorentz structure
$d^{TNP}(q^2; \Delta^2) = d(q^2;
\Delta^2) -  d^{PT}(q^2)$ can be obtained from the relations (2.16)
and (2.14) for $d(q^2; \Delta^2)$ and $d^{PT}(q^2)$, respectively.

The subtraction (2.17) is equivalent to

\begin{equation}
D_{\mu\nu}(q; \Delta^2) = D^{TNP}_{\mu\nu}(q; \Delta^2) +
D^{PT}_{\mu\nu}(q).
\end{equation}
The TNP gluon propagator (2.18) does not survive in the formal PT
$\Delta^2=0$ limit. This means that it is free of the PT
contributions, by construction. The full gluon propagator in this
limit is reduced to its PT counterpart. This means that the full
gluon propagator, being also NP, nevertheless, is "contaminated"
by them. The TNP gluon propagator is purely transversal in a gauge
invariant way (no special (Landau) gauge choice by hand), while
its full counterpart has a longitudinal component as well. There
is no doubt that the true NP dynamics of the full gluon propagator
is completely contained in its TNP part, since the subtraction
(2.19) is nothing but adding zero to the full gluon propagator. We
can write $D_{\mu\nu}(q; \Delta^2) = i \left\{ T_{\mu\nu}(q)
d(q^2; \Delta^2) + \xi L_{\mu\nu}(q) \right\} (1 / q^2) - i
T_{\mu\nu}(q) d^{PT}(q^2)(1 / q^2) + i T_{\mu\nu}(q) d^{PT}(q^2)(1
/ q^2) = D^{TNP}_{\mu\nu}(q; \Delta^2) + D^{PT}_{\mu\nu}(q)$, and
so the true NP dynamics in the full gluon propagator is not
affected, but contrary exactly separated from its PT dynamics,
indeed. In other words, the TNP gluon propagator is the full gluon
propagator but free of its PT "tail".

Taking this important observation into account, we propose instead
of the full gluon propagator to use its TNP counterpart (2.18) as
the relevant gluon propagator for NP QCD, i.e., to replace

\begin{equation}
D_{\mu\nu}(q; \Delta^2)  \rightarrow D^{TNP}_{\mu\nu}(q; \Delta^2)
= D_{\mu\nu}(q; \Delta^2) - D^{PT}_{\mu\nu}(q),
\end{equation}
and hence $d(q^2; \Delta^2) \rightarrow d^{TNP}(q^2; \Delta^2) =
d(q^2; \Delta^2) - d^{PT}(q^2)$.

The subtraction (2.20) plays effectively the role of ghosts in our
proposal. However, the ghosts cancel only the longitudinal
component in the PT gluon propagator, while our proposal leads to
the cancellation of the PT contribution in the full gluon
propagator as well (and thus to an automatical cancellation of its
longitudinal component). Nevertheless, this is not a problem,
since the mass gap is not survived in the formal PT limit, anyway.

In fact, our proposal is reduced to a rather simple prescription.
If one knows a full gluon propagator, and is able to identify the
mass scale parameter responsible for the NP dynamics in it, then
the full gluon propagator should be replaced in accordance with
the subtraction (2.20). The only problem with it is that, being
exact, it may not be unique. However, the uniqueness of such kind
of separation can be achieved only in the explicit solution for
the full gluon propagator as a function of the mass gap (see
below). Anyway, this subtraction is a first necessary step, which
guarantees transversality of the TNP gluon propagator
$D^{TNP}_{\mu\nu}(q; \Delta^2)$ without losing even one bit of
information on the true NP dynamics in the full gluon propagator
$D_{\mu\nu}(q; \Delta^2)$. At the same time, its non-trivial PT
dynamics is completely saved in its PT part $D^{PT}_{\mu\nu}(q)$.
So it is worth emphasizing that the both terms in the subtraction
(2.19) are valid in the whole momentum range, i.e., they are not
asymptotics.

The full gluon propagator (2.19), keeping the mass gap "alive", is
not "physical" in the sense that it cannot be made transversal by
ghosts. Therefore it cannot be used for numerical calculations of
the physical observables from first principles.  However, our
proposal makes it possible to present it as the exact sum of the
two "physical" propagators. The TNP gluon propagator is
automatically transversal, by construction. It fully contains all
the information of the full gluon propagator on its NP context.
Just it should be used in accordance with the prescription (2.20)
in order to calculate the physical observables in low-energy QCD.
In high-energy QCD the PT gluon propagator (2.13) is to be used.
It is free of the mass gap and the ghosts can cancel its
longitudinal component, making it thus transversal ("physical").

Concluding, in this section we have briefly remind how to preserve
the color gauge invariance/symmetry in QCD at non-zero mass gap.
This means that from now on we can forget the relations (2.8) and
(2.9) at all, since there are no any more their negative
consequences for the truly NP QCD. In this connection let us
remind the initial subtraction (2.3) has been done in a gauge
invariant way (i.e., not in a separate propagators, which enter
the skeleton loop integrals, contributing to the full gluon
self-energy).

\section{Massive solution}

One of the direct consequences of the explicit presence of the
mass gap in the full gluon propagator is that the gluon may
acquire an effective mass, indeed \cite{5}. From Eq. (2.16) it
follows that

\begin{equation}
{ 1 \over q^2} d(q^2) = {1 \over q^2 + q^2 \Pi^s(q^2; \xi) +
\Delta^2 c(\xi)},
\end{equation}
where instead of the dependence on $D$ the dependence on $\xi$ is
explicitly shown, while here and below the dependence on all other
parameters is not shown, for simplicity. The full gluon propagator
(2.11) may have a pole-type solution at the finite point if and
only if the denominator in Eq. (3.1) has a zero at this point $q^2
= - m^2_g$ (Euclidean signature), i.e.,

\begin{equation}
- m^2_g  - m^2_g \Pi^s(-m^2_g; \xi) + \Delta^2 c(\xi)=0,
\end{equation}
where $m^2_g \equiv m^2_g(\xi)$ is an effective gluon mass. The
previous equation is a transcendental equation for its
determination. Evidently, the number of its solutions is not
fixed, $a \ priori$. Excluding the mass gap, one obtains that the
denominator in the full gluon propagator becomes

\begin{equation}
q^2 + q^2 \Pi^s(q^2; \xi) + \Delta^2 c(\xi) = q^2 + m^2_g + q^2
\Pi^s(q^2; \xi) + m^2_g \Pi^s(-m^2_g; \xi).
\end{equation}

Let us now expand $\Pi^s(q^2; \xi)$ in a Taylor series near
$m^2_g$:

\begin{equation}
\Pi^s(q^2; \xi) = \Pi^s(-m^2_g; \xi) + (q^2 + m^2_g)
\Pi'^s(-m^2_g; \xi) + O \Bigl( (q^2 + m^2_g)^2 \Bigr).
\end{equation}
Substituting this expansion into the previous relation and after
doing some tedious algebra, one obtains

\begin{equation}
q^2 + m^2_g + q^2 \Pi^s(q^2; \xi) + m^2_g \Pi^s(-m^2_g; \xi)= (q^2
+ m^2_g)[1 +  \Pi^s(-m^2_g; \xi) - m^2_g \Pi'^s(-m^2_g; \xi)] [ 1
+ \Pi^{s,R}(q^2; \xi)],
\end{equation}
where $\Pi^{s,R}(q^2; \xi)= 0$ at $q^2=-m^2_g$ and it is regular
at small $q^2$; otherwise it remains arbitrary.

The full gluon propagator (2.11) thus now looks

\begin{equation}
D_{\mu\nu}(q; m^2_g) = i T_{\mu\nu}(q) {Z_3(m_g^2) \over (q^2 +
m^2_g) [ 1 + \Pi^{s,R}(q^2; m^2_g)]} + i \xi L_{\mu\nu}(q) {1
\over q^2},
\end{equation}
where, for future purpose, in the invariant function
$\Pi^{s,R}(q^2; m^2_g)$ instead of the gauge-fixing parameter
$\xi$ we introduced the dependence on the gluon effective mass
squared $m_g^2$, which depends on $\xi$ itself. The gluon
propagator's renormalization constant is

\begin{equation}
Z_3(m_g^2) = { 1 \over 1 + \Pi^s(-m^2_g; \xi) - m^2_g
\Pi'^s(-m^2_g; \xi)}.
\end{equation}
In the formal PT limit $\Delta^2 =0$, an effective gluon mass is
also zero, $m_g^2(\xi) =0$, as it follows from Eq.~(3.2). So an
effective gluon mass is the NP effect. At the same time, it cannot
be interpreted as the "physical" gluon mass, since it remains
explicitly gauge-dependent quantity (at least at this stage). In
other words, we were unable to renormalize it along with the gluon
propagator (3.6). In the formal PT $\Delta^2 =m_g^2(\xi) =0$ limit
the gluon propagator's renormalization constant (3.7) becomes the
standard one \cite{1,2}, namely

\begin{equation}
Z_3(0) = {1 \over 1 + \Pi^s(0; \xi)}.
\end{equation}

It is interesting to note that Eq. (3.2) has a second solution in
the formal PT $\Delta^2 =0$ limit. In this case an effective gluon
mass remains finite, but $1 + \Pi^s(-m^2_g; \xi) =0$. So a scale
responsible for the NP dynamics is not determined by an effective
gluon mass itself, but by this condition. Its interpretation from
the physical point of view is not clear. The massive solution
(3.6) is difficult to use for the solution of the color
confinement problem, since it is smooth in the $q^2 \rightarrow 0$
limit. However, its existence shows the general possibility for a
vector particles to acquire masses dynamically, i.e., without
so-called Higgs mechanism \cite{6}, which requires the existence of not yet
discovered Higgs particle. Apparently, it can be also useful in the generalization of QCD to non-zero
temperature and density \cite{7,8} (and references therein), when the gluons may indeed acquire effective masses.
The above-mentioned possibility is due only to the internal dynamics and symmetries of the corresponding
gauge theory.

The general procedure described above in subsection E of section
II can be directly applied to the massive solution (3.6). So it
becomes

\begin{equation}
D_{\mu\nu}(q; m^2_g) =   D^{TNP}_{\mu\nu}(q; m^2_g) +
D^{PT}_{\mu\nu}(q),
\end{equation}
where

\begin{equation}
D^{TNP}_{\mu\nu}(q; m^2_g) = i T_{\mu\nu}(q) \left[ {Z_3(m^2_g)
\over (q^2 + m^2_g) [ 1 + \Pi^{s,R}(q^2; m^2_g)]} - {Z_3(0) \over
q^2 [ 1 + \Pi^{s,R}(q^2; 0)]} \right]
\end{equation}
and

\begin{equation}
D^{PT}_{\mu\nu}(q) = i \left[ T_{\mu\nu}(q) {Z_3(0) \over [ 1 +
\Pi^{s,R}(q^2; 0)]} + \xi L_{\mu\nu}(q) \right] {1 \over q^2}.
\end{equation}
Let us remind that in the massive solution the role of the mass
gap is played by an effective gluon mass, so the formal PT limit
is $m^2_g=0$. In accordance with our prescription (2.20), we
should finally replace the full gluon propagator (3.6) as follows:
$D_{\mu\nu}(q; m_g^2) \rightarrow D^{TNP}_{\mu\nu}(q; m_g^2)$,
where the latter is explicitly given in Eq.~(3.10).

\section{General NL iteration solution}

In order to find another type of the general formal solution for the
full gluon propagator (2.15), let us begin again with
its "solution" (2.16) which is

\begin{equation}
d(q^2) \equiv d(q^2; \Delta^2) = {1 \over 1 + \Pi^s(q^2; d) + c(d) (\Delta^2 / q^2) },
\end{equation}
where the dependence on $D$ is replaced by the equivalent
dependence on $d$ and the relation (2.4) is already used. It is worth reminding that the invariant function
$\Pi^s(q^2; d)$ and $c(d)$ are, in fact, the sum of the corresponding skeleton loop integrals (see section II and
our initial paper \cite{4}). Let us introduce further the dimensionless variable
$z= \Delta^2 / q^2$. The full Lorentz structure (4.1) regularly depends on the mass gap, and hence on $z$.
Thus it can be expand in a Taylor series in powers of $z$ around zero $z$ as follows:

\begin{equation}
d(q^2; \Delta^2) = d(q^2; z)= \sum_{k=0}^{\infty} z^k f_k(q^2),
\end{equation}
where the functions $f_k(q^2)$ are the corresponding derivatives of $d(q^2; z)$ with respect to $z$ at $z=0$,
which is equivalent to the PT $\Delta^2 =0$ limit. For example, $f_0(q^2) = d(q^2; z=0) = d^{PT}(q^2) = [1+
\Pi^s(q^2; d^{PT}]^{-1}$, $f_1(q^2) = (\partial d(q^2; z) / \partial z)_{z=0} =
\left[ \partial [1 + \Pi^s(q^2; d) + c(d)z]^{-1} / \partial z
\right]_{z=0} = - \left[ 1 + \Pi^s(q^2; d^{PT}) \right]^{-2}
c(d^{PT})= - [d^{PT}(q^2)]^2 c(d^{PT})$, and so on, i.e.,
$f_k(q^2)= (-1)^k  d^{PT}(q^2) [d^{PT}(q^2) c(d^{PT})]^k$. Fortunately, these
explicit expressions play no any role in what follows. In any
case, they depend on the unknown, in general, quantities
$\Pi^s(q^2; d)$ and $c(d)$, which by themselves NL depend on $d$ and finally on $d^{PT}$ and $c(d^{PT})$.
So our expansion (4.2) is nothing but the NL iteration series in
powers of the mass gap (for the direct NL iteration procedure
with $d^{(0)}=1$ as input information see appendix A). To use also unknown functions
$f_k(q^2)$ much more convenient from the technical point of view.
However, it is worth emphasizing that, contrary to the relation (4.1), the expansion (4.2) can be considered
now as a formal solution for $d(q^2)$, since $f_k(q^2)$ depend on $d^{PT}(q^2)$, which is assumed to be
"known".

The functions $f_k(q^2)$ are regular functions of the
variable $q^2$, since they finally depend on $d^{PT}(q^2)$ which
is a regular function of $q^2$. Therefore they can be
expand in a Taylor series near $q^2=0$ (here we can put the
subtraction point $\alpha=0$, for simplicity, since all the
quantities are already regularized, i.e., they depend on $\alpha$
and so on, see appendix A). Introducing the dimensionless
variable $x = q^2 / M^2$, where $M^2$ is some fixed auxiliary mass squared,
it is convenient to present this expansion as a sum of the two
terms, namely

\begin{equation}
f_k(q^2) = \sum_{n=0}^k x^n f_{kn}(0) + x^{k+1}B_k(x),
\end{equation}
where the coefficient $f_{kn}(0)$ are the corresponding
derivatives of the functions $f_k(q^2) \equiv
f_k(x)$ with respect to $x$ at $x=0$. Of course, these
coefficients depend on the parameters of the theory such as
$\lambda, \alpha, \xi, g^2$, and so on, which are not shown
explicitly. The dependence on these parameters will be restored at
the final stage of our derivations. The dimensionless functions
$B_k(x)$ are regular functions of $x$; otherwise they remain
arbitrary.

So the general Lorentz structure (4.2) becomes

\begin{equation}
d(q^2) =  \sum_{k=0}^{\infty} z^k f_k(x) =
\sum_{k=0}^{\infty} z^k \Big( \sum_{n=0}^k x^n
f_{kn}(0) + x^{k+1} B_k(x) \Big).
\end{equation}
Omitting all the intermediate tedious derivations (which,
nevertheless, are quite obvious), these double sums can be
equivalently present as the sum of the three independent terms as
follows:

\begin{equation}
d(q^2) =  z \sum_{k=0}^{\infty} z^k \sum_{m=0}^{\infty}
\Phi_{km}(0) + a \sum_{k=0}^{\infty} a^k
\sum_{m=0}^{\infty}A_{km}(x) + d^{PT}(q^2),
\end{equation}
where the constant $a = xz = \Delta^2 / M^2$ and the
dimensionless functions $A_{km}(x)$ are regular functions of $x$:
otherwise they remain arbitrary. $d^{PT}(q^2)$ denotes the terms which do not depend on the mass
gap $\Delta^2$ at all, i.e., it is nothing but the Lorentz
structure of the PT gluon propagator (2.14), indeed. The summation over $m$ explicitly shows that all
iterations invoke each NP IR singularity labeled by $k$ in the
first term of the expansion (4.5).  Thus it is the general NL formal expansion in powers of the
mass gap (this is explicitly seen from appendix A).

Going back to the gluon momentum variable $q^2$, one obtains

\begin{equation}
d(q^2; \Delta^2) = d^{TNP}(q^2; \Delta^2)
+ d^{PT}(q^2) = d^{INP}(q^2; \Delta^2) + d^{MPT}(q^2; \Delta^2)
+ d^{PT}(q^2),
\end{equation}
where the superscripts "INP" and "MPT" stand for
the intrinsically NP and mixed PT parts of the TNP term,
respectively (for reasons see discussion below). In other words, in the general NL iteration solution the
TNP part itself is a sum of the two independent terms, i.e., $d^{TNP}(q^2; \Delta^2) = d^{INP}(q^2; \Delta^2) +d^{MPT}(q^2; \Delta^2)$. Their explicit expressions are

\begin{equation}
d^{INP}(q^2; \Delta^2) = \Bigl( {\Delta^2 \over q^2} \Bigr)
\sum_{k=0}^{\infty} \Bigl( {\Delta^2 \over q^2} \Bigr)^k \Phi_k =
\Bigl( {\Delta^2 \over q^2} \Bigr) \sum_{k=0}^{\infty} \Bigl(
{\Delta^2 \over q^2} \Bigr)^k \sum_{m=0}^{\infty} \Phi_{km}
\end{equation}
and

\begin{equation}
d^{MPT}(q^2; \Delta^2) = \Bigl( {\Delta^2 \over M^2} \Bigr)
\sum_{k=0}^{\infty} \Bigl( {\Delta^2 \over M^2} \Bigr)^k A_k(q^2)
= \Bigl( {\Delta^2 \over M^2} \Bigr) \sum_{k=0}^{\infty} \Bigl(
{\Delta^2 \over M^2} \Bigr)^k \sum_{m=0}^{\infty} A_{km}(q^2).
\end{equation}
Here and  everywhere below all the quantities depend on the
parameters of the theory, namely $\Delta^2 = \Delta^2(\lambda,
\alpha, \xi, g^2)$ and $A_k(q^2) = \sum_{m=0}^{\infty} A_{km}(q^2;
\lambda, \alpha, \xi, g^2)$. At the same time, $\Phi_{km}$ depends
in addition on the parameter $a$ as well, i.e., $\Phi_{km} =
\Phi_{km}(\lambda, \alpha, \xi, g^2, a)$.

\subsection{The exact structure of the NL iteration solution}

The full gluon propagator (2.11) thus
becomes the sum of the three independent terms, namely

\begin{equation}
D_{\mu\nu}(q; \Delta^2) = D^{TNP}_{\mu\nu}(q; \Delta^2)+
D^{PT}_{\mu\nu}(q) = D^{INP}_{\mu\nu}(q; \Delta^2)+
D^{MPT}_{\mu\nu}(q; \Delta^2) + D^{PT}_{\mu\nu}(q),
\end{equation}
where

\begin{equation}
D^{INP}_{\mu\nu}(q; \Delta^2) = i T_{\mu\nu}(q) d^{INP}(q^2;
\Delta^2) {1 \over q^2} = i T_{\mu\nu}(q) {\Delta^2 \over (q^2)^2}
L (q^2; \Delta^2)
\end{equation}
with

\begin{equation}
L(q^2; \Delta^2) = \sum_{k=0}^{\infty} \Bigl( {\Delta^2 \over q^2}
\Bigr)^k \Phi_k = \sum_{k=0}^{\infty} \Bigl( {\Delta^2 \over q^2}
\Bigr)^k \sum_{m=0}^{\infty} \Phi_{km},
\end{equation}
while

\begin{equation}
D^{MPT}_{\mu\nu}(q; \Delta^2) = i T_{\mu\nu}(q) d^{MPT}(q^2;
\Delta^2) {1 \over q^2}
\end{equation}
with $d^{MPT}(q^2; \Delta^2)$ given in Eq.~(4.13) and

\begin{equation}
D^{PT}_{\mu\nu}(q) = i \Bigr[ T_{\mu\nu}(q) d^{PT}(q^2) + \xi
L_{\mu\nu}(q) \Bigl] {1 \over q^2}
\end{equation}
with $d^{PT}(q^2)$ given in Eq.~(2.14). For the direct NL
iteration procedure see appendix A, as mentioned above.

Let us emphasize that the general problem of convergence of formal
(but regularized) series, which appear in these relations, is
irrelevant here. In other words, it does not make any sense to
discuss the convergence of such kind of series before the
renormalization program is performed (which will allow one to see
whether or not the mass gap survives it at all). The problem how
to remove the UV overlapping divergences \cite{9} and usual
overall ones \cite{1,2,10,11} is a standard one, i.e., it is not our
problem, anyway (let us remind that the mass gap does not survive
in the PT $q^2 \rightarrow \infty$ limit). Our problem will be how
to deal with severe infrared (IR) ($q^2 \rightarrow 0$)
singularities due to their novelty and genuine (intrinsic) NP
character (in this limit the mass gap dominates the structure of
the full gluon propagator). Fortunately, there already exists a
well-elaborated mathematical formalism for this purpose, namely
the distribution theory (DT) \cite{12}, into which the dimensional
regularization method (DRM) \cite{13} should be correctly
implemented (see also Refs. \cite{14,15}).

The INP part of the full gluon propagator is characterized by the
presence of severe power-type (or, equivalently, NP) IR
singularities $(q^2)^{-2-k}, \ k=0,1,2,3,...$. So these IR
singularities are defined as more singular than the power-type IR
singularity of the free gluon propagator $(q^2)^{-1}$, which thus
can be defined as the PT IR singularity. The INP part of the full
gluon propagator (4.10), apart from the structure $(\Delta^2 /
q^4)$, is nothing but the corresponding Laurent expansion
(explicitly shown in Eq. (4.11)) in integer powers of $q^2$
accompanied by the corresponding powers of the mass gap squared
and multiplied by the $q^2$-independent factors, the so-called
residues $\Phi_k(\lambda, \alpha, \xi, g^2, a) =
\sum_{m=0}^{\infty} \Phi_{km}(\lambda, \alpha, \xi, g^2, a)$. The
sum over $m$ indicates that an infinite number of iterations (all
iterations) of the above-mentioned corresponding regularized
skeleton loop integrals invokes each severe IR singularity
labeled by $k$. It is worth emphasizing that the Laurent
expansion (4.11) cannot be summed up into the some known function,
since its residues are, in general, arbitrary. However, this
arbitrariness is not a problem. The functional dependence, which
has been established exactly, is all that matters (this will be
explicitly shown in the subsequent paper). Let us note that the
expansions (4.10)-(4.11) have been independently obtained in Ref.
\cite{14} in a rather different way.

The MPT part of the full gluon propagator (4.12), which has the
power-type PT IR singularity only, remains undetermined, but
depends on the mass gap (that is why we call this term as the
mixed PT contribution, but it vanishes in the formal PT
$\Delta^2=0$ limit). This is the price we have paid to fix exactly
the functional dependence of the INP part of the full gluon
propagator. With respect to the character of the IR singularity it
should be combined with the PT gluon propagator, leading to the so-called
general PT (GPT) term, namely

\begin{equation}
D^{GPT}_{\mu\nu}(q; \Delta^2) = D^{MPT}_{\mu\nu}(q; \Delta^2) + D^{PT}_{\mu\nu}(q) =
i \Bigr[ T_{\mu\nu}(q) d^{GPT}(q^2; \Delta^2) + \xi
L_{\mu\nu}(q) \Bigl] {1 \over q^2},
\end{equation}
where $d^{GPT}(q^2; \Delta^2) = d^{MPT}(q^2; \Delta^2) +  d^{PT}(q^2)$ is regular at small $q^2$, while
$d^{MPT}(q^2; \Delta^2=0)= 0$ and hence $d^{GPT}(q^2; \Delta^2=0)= d^{PT}(q^2)$. Thus both terms
MPT and PT present the PT-type contributions to the full gluon propagator (4.6).
It is worth reminding that all the three terms, which
appear in the right-hand-side of Eq.~(4.9) are valid in the whole
energy/momentum range, i.e., they are not asymptotics. At the same
time, we have achieved the separation between the terms
responsible for the NP (dominating in the IR ($q^2 \rightarrow
0$)) and the nontrivial PT (dominating in the UV ($q^2 \rightarrow
\infty$)) dynamics in the true QCD vacuum. The structure of this
solution shows clearly that the deep IR region interesting for
confinement and other NP effects is dominated by the mass gap. In
the formal PT $\Delta^2=0$ limit, the nontrivial PT dynamics is
all that matters.

\section{INP gluon propagator}

In accordance with our prescription, one should subtract all the
types of the PT contributions in order to get the relevant gluon
propagator for the truly NP QCD. As it follows from discussion above, in the case of the
NL iteration solution, we should subtract the two terms. Doing so
in Eq.~(4.9), on account of Eq.~(4.14), one finally obtains

\begin{equation}
D_{\mu\nu}(q; \Delta^2) \rightarrow  D^{INP}_{\mu\nu}(q; \Delta^2)
= D_{\mu\nu}(q; \Delta^2) - D^{GPT}_{\mu\nu}(q; \Delta^2),
\end{equation}
and hence $d(q^2) \rightarrow d^{INP}(q^2)$ as well, so that

\begin{equation}
D^{INP}_{\mu\nu}(q; \Delta^2) = i T_{\mu\nu}(q) {\Delta^2 \over
(q^2)^2} L (q^2; \Delta^2)= i T_{\mu\nu}(q) {\Delta^2 \over
(q^2)^2} \sum_{k=0}^{\infty} \Bigl( {\Delta^2 \over q^2} \Bigr)^k
\Phi_k,
\end{equation}
where $\Delta^2 = \Delta^2(\lambda, \alpha, \xi, g^2)$ and $\Phi_k
= \Phi_k(\lambda, \alpha, \xi, g^2) = \sum_{m=0}^{\infty}
\Phi_{km}(\lambda, \alpha, \xi, g^2)$. In this connection, let us
note that after the subtraction (5.1) is completed we can put the
intermediate parameter $a = 1$, to equate thus the auxiliary fixed
mass to the mass gap itself, i.e., put $M^2 = \Delta^2$, not
losing generality. In the deep IR region ($q^2 \rightarrow 0$) the
mass gap is only one that's really matters. All other masses
introduced from a technical point of view in order to clarify the
derivations play only auxiliary role.

It is important to emphasize that the INP gluon propagator (5.2)
is {\bf uniquely} defined because there exists a special
regularization expansion for severe (i.e., NP) IR singularities,
while for the PT IR singularity such kind of expansion does not
exist at all (see Refs. \cite{12,14,15} and references therein).
This just determines the principal difference between the NP and
PT IR singularities. It is also {\bf exactly} defined because of
its two features. The first one is that the INP gluon propagator
depends only on the transversal degrees of freedom of gauge
bosons. The second one is that in the formal PT $\Delta^2 =0$
limit the INP gluon propagator vanishes. Thus, one can conclude
that the presence of severe IR singularities only is the first
necessary condition, while the regular dependence on the mass gap
and transversality is only second sufficient condition for the
unique and exact separation of the INP gluon propagator from the
PT gluon propagator. At the same time, the TNP gluon propagator is
not uniquely defined, since it contains the MPT part, see Eq.~(4.9).
In other words, the INP gluon propagator is free of all
the types of the PT contributions ("contaminations"). Just it
should replace the full gluon propagator in order to calculate the
physical observables, processes, etc. from first principles in
low-energy QCD after the corresponding renormalization program is
performed.

The INP gluon propagator satisfies its own equation of motion. For
the sake of completeness, let us begin with the SD equation for
the TNP gluon propagator \cite{4}, namely

\begin{eqnarray}
D^{TNP}_{\mu\nu}(q; \Delta^2) &=& D^0_{\mu\rho}(q)i
T_{\rho\sigma}(q)[ - q^2 \Pi^s(q^2; D^{PT}) + q^2 \Pi^s(q^2; D)
+ \Delta^2] D^{PT}_{\sigma\nu}(q) \nonumber\\
&+& D^0_{\mu\rho}(q)i T_{\rho\sigma}(q) [q^2 \Pi^s(q^2; D) +
\Delta^2] D^{TNP}_{\sigma\nu}(q; \Delta^2)
\end{eqnarray}
with

\begin{equation}
D^{TNP}_{\mu\nu}(q) = i \left\{ T_{\mu\nu}(q) d^{TNP}(q^2) + \xi
L_{\mu\nu}(q) \right\} {1 \over q^2}.
\end{equation}
Here and below we omit the dependence on the mass gap in the
propagators and their Lorentz structures, for simplicity. On
account of this decomposition, the "solution" of the previous
equation is

\begin{equation}
d^{TNP}(q^2) = { \Pi^s(q^2; D^{PT}) - \Pi^s(q^2; D) - ( \Delta^2 /
q^2) \over [ 1 + \Pi^s(q^2; D) + (\Delta^2 / q^2) ] [ 1 +
\Pi^s(q^2; D^{PT}) ] }.
\end{equation}
This expression coincides with the definition of $d^{TNP}(q^2) =
d(q^2) - d^{PT}(q^2)$ on account of the explicit expressions
(2.14) and (2.15), as it should be.

From Eq.~(4.9) it follows that

\begin{equation}
D^{TNP}_{\mu\nu}(q) = D^{INP}_{\mu\nu}(q)+ D^{MPT}_{\mu\nu}(q),
\end{equation}
and substituting it into Eq.~(5.3), one obtains the SD equation
for the INP gluon propagator, namely

\begin{eqnarray}
D^{INP}_{\mu\nu}(q) = &-& D^{MPT}_{\mu\nu}(q) + D^0_{\mu\rho}(q)i
T_{\rho\sigma}(q) [q^2 \Pi^s(q^2; D) + \Delta^2]
D^{MPT}_{\sigma\nu}(q) \nonumber\\
&+& D^0_{\mu\rho}(q)i T_{\rho\sigma}(q)[ - q^2 \Pi^s(q^2; D^{PT})
+ q^2 \Pi^s(q^2; D) + \Delta^2] D^{PT}_{\sigma\nu}(q) \nonumber\\
&+& D^0_{\mu\rho}(q)i T_{\rho\sigma}(q) [q^2 \Pi^s(q^2; D) +
\Delta^2] D^{INP}_{\sigma\nu}(q).
\end{eqnarray}
Using the decompositions (2.2), (4.12) and (4.13), it can be
simplified to

\begin{eqnarray}
q^2 D^{INP}_{\mu\nu}(q) = &-& i T_{\mu\nu}(q) \Bigl( 1 +
\Pi^s(q^2; D) + (\Delta^2 / q^2) \Bigr) d^{MPT}(q) \nonumber\\
&-& i T_{\mu\nu}(q) \Bigl( - \Pi^s(q^2; D^{PT}) +
\Pi^s(q^2; D) + (\Delta^2 / q^2) \Bigr) d^{PT}(q)  \nonumber\\
&-& T_{\mu\sigma}(q) T_{\rho\sigma}(q) [q^2 \Pi^s(q^2; D) +
\Delta^2] D^{INP}_{\sigma\nu}(q),
\end{eqnarray}
where $d^{MPT}(q)$ and $d^{PT}(q)$ are given in Eqs.~(4.8) and
(2.14), respectively. This equation is of no practical
use due to its complicated structure. Fortunately, we already have
the explicit expression for the INP gluon propagator (4.10)-(4.11)
or, equivalently, (5.2). It is only one to be used in order to
derive renormalized gluon propagator with the correct confinement
properties.

However, from Eq.~(5.8) it follows an important observation that
like the TNP SD equation (5.3) this equation cannot be reduced to
the free gluon propagator, when the interaction is to be switched
off (i.e., setting formally $\Pi^s(q^2; D^{PT}) = \Pi^s(q^2; D) =
\Delta^2 =0$). Evidently, to the same conclusion one comes from the explicit expressions
(4.8) and (5.5), on account of the relation
$d^{INP}(q^2; \Delta^2) = d^{TNP}(q^2; \Delta^2) - d^{MPT}(q^2; \Delta^2)$,
which follows from Eq.~(4.6). So in INP QCD the gluon propagator is always
"dressed" as well, and thus this theory has no free gluon
propagator in its formalism. As it has been argued in our initial
work \cite{4}, it makes it possible to suppress the emission and
absorbtion of the colored dressed gluons at large distances by the
renormalization of the mass gap. Both the suppression of the
dressed gluons and the absence of the free gluons are necessary
for the explanation of gluon confinement by INP QCD (see our next
paper). On the other hand, the full gluon propagator (2.19) which
satisfies Eq.~(2.15) is reduced to the free gluon propagator when
the interaction is switched off. There is no mechanism to suppress
the emission and absorbtion of the free gluons at large distances \cite{4}. That is
why the full gluon propagator (2.19) is not confining, while the
INP one (5.2) can be.

{\bf The subtraction (5.1) seems to be necessary, indeed. It makes
the relevant gluon propagator (5.2) transversal and excludes the free
gluons from the theory at the same time}.

\section{Conclusions}

The structure of the full gluon propagator in the presence of the
regularized mass gap has been firmly established. We have shown
explicitly that in its presence at least two independent and
different typed of formal solutions for the regularized full gluon
propagator exist. No truncations/approximations/assumptions are
made in order to show the existence of these general types of
solutions. Also, our approach, in general, and the above-mentioned
solutions, in particular, is gauge-invariant, since no special
gauge has been chosen. Let us emphasize that before the
renormalization program is performed the gauge invariance should
be understood in this sense only.

In the presence of the mass gap the gluons may acquire an
effective gluon masses, depending on the gauge choice (the
so-called massive solution (3.6)), but a gauge-fixing parameter
remains arbitrary, i.e., a gauge is not fixed by hand (see remarks
above). Its relation to the solution of the color confinement
problem is not clear, even after the renormalization program is
performed.

The general NL iteration solution (4.9)-(4.13) for the full gluon
propagator depends explicitly on the mass gap. It is always
severely singular in the $q^2 \rightarrow 0$ limit, so the gluons
remain massless, and this does not depend on the gauge choice.
However, we argued that only the INP gluon propagator (5.2) is to
be used for the numerical calculations of physical observables,
processes, etc. in low-energy QCD from first principles.

It is worth emphasizing that there exists only one general
restriction on the behavior of $\Pi^s(q^2; D)$, which enters the
corresponding gluon SD equation (2.15), in the explicit presence
of the mass gap within our approach, namely

\begin{equation}
q^2 \Pi^s(q^2; D) \rightarrow 0, \quad q^2 \rightarrow 0,
\end{equation}
at any $D$. It stems from the second of the exact decompositions
(2.5), since the subtracted gluon self-energy in this limit (or
more precisely at $q^2 \rightarrow \mu^2$) should go to zero.
Otherwise the invariant function $\Pi^s(q^2; D)$ remains arbitrary
(but it is logarithmic divergent at infinity). Both general types
of formal solutions the massive solution and the NL iteration one satisfy
it. The existence of some other solution(s) for the full gluon
propagator, satisfying the general condition (6.1), should not be
excluded $a \ priori$. Let us remind that the gluon SD equation
(2.15) is highly NL, so the number of independent solutions is
not fixed. Any concrete solution obtained by lattice QCD or by the analytical approach based on the SD system of equations is a particular case of the general types (finite or singular at zero gluon momentum) of the formal solutions established here. They are subject to the different truncations/approximations/assumptions and the concrete gauge choice imposed on the invariant function $\Pi^s(q^2; D)$, which, in general, remains arbitrary but satisfying
the above-mentioned general constraint (6.1) within our approach (see, for example recent papers \cite{16,17,18,19,20,21,22} and references therein. Let us also point out Refs. \cite{23,24,25,26} as well, where the
gluon propagator is finite and contains the mass scale parameter. However, it, apparently, cannot be interpreted as 
gluon effective mass).

The INP solution (5.2) is interesting for confinement, but the two
important problems remain to solve. The first problem is how to
perform the renormalization program for the regularized mass gap
$\Delta^2 \equiv \Delta^2(\lambda, \alpha, \xi, g^2)$, and to see
whether the mass gap survives it or not (it has been already
discussed in our previous work \cite{4}). The second problem is
how to treat correctly severe IR singularities $(q^2)^{-2-k}, \
k=0,1,2,3,...$ inevitably present in this solution (see a few
brief remarks above in section IV). Both problems will be
addressed and solved in our subsequent paper.

\begin{acknowledgments}

Support by HAS-JINR grant (P. Levai) is to be acknowledged. The
author is grateful to P. Forg\'{a}cs, J. Nyiri, C. Wilkin, T. Bir\'{o}, M. Faber,
\'{A}. Luk\'{a}cs, M. Vas\'{u}th and especially to A.V. Kouzushin for useful discussions, remarks and
help.

\end{acknowledgments}

\appendix

\section{Direct NL iteration procedure}

In order to find a formal solution for the regularized full gluon
propagator (2.11), on account of its effective charge (2.16), let
us rewrite the latter one in the form of the corresponding
transcendental (i.e., not algebraic) equation, namely

\begin{equation}
d(q^2) = 1 - \Bigl[ \Pi^s(q^2; d) + {\Delta^2 \over q^2} c(d)
\Bigr] d(q^2) = 1 - P(q^2; d) d(q^2),
\end{equation}
where Eq.~(2.4) has been already used, and instead of $D$ an
equivalent dependence on $d$ is introduced. It is suitable for the
formal NL iteration procedure. For future purposes, it is
convenient to introduce short-hand notations as follows:

\begin{eqnarray}
c(d=d^{(0)} + d^{(1)} + d^{(2)} + ... + d^{(m)}+ ... ) &=& c_m \equiv c_m(\lambda,
\alpha, \xi, g^2), \nonumber\\
\Pi^s(q^2; d=d^{(0)} + d^{(1)}+d^{(2)}+ ... + d^{(m)} + ...) &=&
\Pi^s_m(q^2),
\end{eqnarray}
and

\begin{equation}
P_m(q^2) = \Bigl[ \Pi^s_m(q^2) + {\Delta^2 \over q^2} c_m \Bigr],
\ m=0,1,2,3,... \ .
\end{equation}
Via the corresponding subscript $m$ it is explicitly seen which
iteration for the gluon form factor $d$ is actually done in
$c(d)$, $\Pi^s(q^2; d)$ and $P(q^2; d)$. Let us also point out
that all the invariant functions $\Pi^s_m(q^2)$ can be expand in a
formal Taylor series near the finite subtraction point $\alpha$.
If it were possible to express the full gluon form factor $d(q^2)$
in terms of these quantities then it would be the formal solution
for the full gluon propagator. In fact, this is nothing but the
skeleton loops expansion, since the regularized skeleton loop
integrals, contributing to the gluon self-energy as mentioned
above, have to be iterated. This is the so-called general NL
iteration solution. This formal expansion is not a PT series. The
magnitude of the coupling constant squared and the dependence of
the regularized skeleton loop integrals on it is completely
arbitrary.

It is instructive to describe the general iteration procedure in
some details. Evidently, $d^{(0)}=1$, and this corresponds to the
approximation of the full gluon propagator by its free
counterpart. Doing the first iteration in Eq. (A1), one thus
obtains

\begin{equation}
d(q^2) = 1 - P_0(q^2) + ... = 1 + d^{(1)}(q^2) + ...,
\end{equation}
where obviously

\begin{equation}
d^{(1)}(q^2) = - P_0(q^2).
\end{equation}
Carrying out the second iteration, one gets

\begin{equation}
d(q^2) = 1 - P_1(q^2) [ 1 + d^{(1)}(q^2) ] + ... = 1 +
d^{(1)}(q^2) + d^{(2)}(q^2) + ...,
\end{equation}
where

\begin{equation}
d^{(2)}(q^2) = - d^{(1)}(q^2) - P_1(q^2) [ 1 - P_0(q^2)].
\end{equation}
Doing the third iteration, one further obtains

\begin{equation}
d(q^2) = 1 - P_2(q^2) [ 1 + d^{(1)}(q^2) + d^{(2)}(q^2)] + ... = 1
+ d^{(1)}(q^2) + d^{(2)}(q^2) + d^{(3)}(q^2) + ...,
\end{equation}
where

\begin{equation}
d^{(3)}(q^2) = - d^{(1)}(q^2) - d^{(2)}(q^2) - P_2(q^2) [ 1 -
P_1(q^2)(1 - P_0(q^2))],
\end{equation}
and so on for the next iterations.

Thus up to the third iteration, one finally arrives at

\begin{equation}
d(q^2) = \sum_{m=0}^{\infty} d^{(m)}(q^2) = 1 - \Bigl[
\Pi^s_2(q^2) + {\Delta^2 \over q^2} c_2 \Bigr] \Bigl[ 1 - \Bigl[
\Pi^s_1(q^2) + {\Delta^2 \over q^2}c_1 \Bigr] \Bigl[ 1 -
\Pi^s_0(q^2) - {\Delta^2 \over q^2} c_0 \Bigr] \Bigr] + ... \ .
\end{equation}
We restrict ourselves by the iterated gluon form factor up to the
third term, since this already allows to show explicitly some
general features of the NL iteration solution.

\subsection{Splitting/shifting procedure}

Doing some tedious algebra, the previous expression (A10) can be
rewritten as follows:

\begin{eqnarray}
d(q^2) &=& \Bigl[ 1 - \Pi^s_2(q^2) + \Pi^s_1(q^2) \Pi^s_2(q^2) -
\Pi^s_0(q^2)
\Pi^s_1(q^2)\Pi^s_2(q^2) + ... \Bigr] \nonumber\\
&+& {\Delta^2 \over q^2} \Bigl[ \Pi^s_2(q^2) c_1 + \Pi^s_1(q^2)
c_2 - \Pi^s_0(q^2) \Pi^s_1(q^2) c_2 - \Pi^s_0(q^2) \Pi^s_2(q^2)
c_1 - \Pi^s_1(q^2) \Pi^s_2(q^2) c_0 + ... \Bigr] \nonumber\\
&-& {\Delta^4 \over q^4} \Bigl[ \Pi^s_0(q^2) c_1 c_2 +
\Pi^s_1(q^2) c_0 c_2 + \Pi^s_2(q^2) c_0 c_1 + ... \Bigr] \nonumber\\
&-& {\Delta^2 \over q^2} \Bigl[ c_2 -  {\Delta^2 \over q^2}c_1 c_2
+ { \Delta^4 \over q^4} c_0 c_1 c_2 + ... \Bigr].
\end{eqnarray}
This formal expansion contains three different types of terms. The
first type are the terms which contain only different combinations
of $\Pi^s_m(q^2)$ (they are not multiplied by inverse powers of
$q^2$); the third type of terms contains only different
combinations of $(\Delta^2 / q^2)$. The second type of terms
contains the so-called mixed terms, containing the first and third
types of terms in different combinations. The two last types of
terms are multiplied by the corresponding powers of $1/q^2$. Such
structure of terms will be present in each iteration term for the
full gluon form factor. However, any of the mixed terms can be
split exactly into the first and third types of terms. For this
purpose the formal Taylor expansions for $\Pi^s_m(q^2)$ around the
finite subtraction point $\alpha$ should be used. Thus an exact IR
structure of the full gluon form factor (which just is our primary
goal to establish) is determined not only by the third type of
terms. It gains contributions from the mixed terms as well, but
without changing its functional dependence (see remarks below). To
demonstrate this in some detail, it is convenient to express the
previous expansion (A11) in terms of dimensionless variables and
parameters introduced in section IV, namely

\begin{equation}
z = {\Delta^2 \over q^2}, \quad x = {q^2 \over M^2}, \quad a = zx=
{\Delta^2 \over M^2}, \quad \alpha = {\mu^2 \over M^2},
\end{equation}
where $M^2$ is some fixed mass squared, and $\mu^2$ is the fixed
point close to $q^2=0$ (to be not mixed up with the tensor index).
Also, in the formal PT $\Delta^2 =0$ limit $a=0$ as well, since
$M^2$ is fixed. On account of the relations (A12), the expansion
(A11) becomes

\begin{eqnarray}
d(x) &=& \Bigl[ 1 - \Pi^s_2(x) + \Pi^s_1(x) \Pi^s_2(x) -
\Pi^s_0(x)
\Pi^s_1(x)\Pi^s_2(x) + ... \Bigr] \nonumber\\
&+& z \Bigl[ \Pi^s_2(x) c_1  + \Pi^s_1(x)c_2 - \Pi^s_0(x)
\Pi^s_1(x) c_2  - \Pi^s_0(x) \Pi^s_2(x) c_1
- \Pi^s_1(x) \Pi^s_2(x) c_0  + ... \Bigr] \nonumber\\
&-& z^2 \Bigl[ \Pi^s_0(x) c_1 c_2  + \Pi^s_1(x) c_0
c_2 + \Pi^s_2(x) c_0 c_1  + ... \Bigr] \nonumber\\
&-& z \Bigl[ c_2  - \Bigl( { a \over x} \Bigr) c_1 c_2 + \Bigl( {a
\over x} \Bigr)^2 c_0 c_1 c_2 + ... \Bigr].
\end{eqnarray}

Taking into account the above-mentioned formal Taylor expansions

\begin{equation}
\Pi^s_m(x) = \sum_{n=0}^{\infty} (x - \alpha)^n \Pi^{(n)}_m
(\alpha) = \sum_{n=0}^{\infty} \Bigl[ \sum_{k=0}^n p_{nk} x^k
\alpha^{n-k} \Bigr] \Pi^{(n)}_m (\alpha),
\end{equation}
for example, the mixed term $z \Pi^s_2(x) c_1$ can be then exactly
split/decomposed as follows:

\begin{equation}
c_1 z \Pi^s_2(x) = c_1 z \sum_{n=0}^{\infty} \Bigl[ \sum_{k=0}^n
p_{nk} x^k \alpha^{n-k} \Bigr] \Pi^{(n)}_2 (\alpha) = z
P_1(\alpha) + P_0(\alpha) + O_2(x).
\end{equation}
Here and below the dependence on all other possible parameters is
not shown, for simplicity. The dimensionless function $O_2(x)$ is
of the order $x$ at small $x$; otherwise it remains arbitrary. The
first term now is to be shifted to the third type of terms, while
the remaining terms are to be shifted to the first type of terms.
All other mixed terms of similar structure should be treated
absolutely in the same way.

The mixed term $z^2 \Pi^s_0(x)c_1 c_2$ can be split as

\begin{equation}
c_1 c_2 z^2 \Pi^s_0(x) = c_1 c_2 z^2 \sum_{n=0}^{\infty} \Bigl[
\sum_{k=0}^n p_{nk} x^k \alpha^{n-k} \Bigr] \Pi^{(n)}_0 (\alpha) =
z^2 P_2(\alpha) + z N_1(\alpha) + N_0(\alpha) + O_0(x),
\end{equation}
where the dimensionless function $O_0(x)$ is of the order $x$ at
small $x$; otherwise it remains arbitrary. Again the first two
terms should be shifted to the third type of terms, while the last
two terms should be shifted to the first type of terms.

Similarly to the formal Taylor expansion (A14), we can write

\begin{equation}
\Pi^s_m(x)\Pi^s_{m'} (x)= \Pi^s_{mm'} (x) = \sum_{n=0}^{\infty} (x
- \alpha)^n \Pi^{(n)}_{mm'} (\alpha) = \sum_{n=0}^{\infty} \Bigl[
\sum_{k=0}^n p_{nk} x^k \alpha^{n-k} \Bigr] \Pi^{(n)}_{mm'}
(\alpha).
\end{equation}
Then, for example the mixed term $z \Pi^s_0(x) \Pi^s_1(qx) c_2$
can be split as

\begin{equation}
c_2 z \Pi^s_0(x) \Pi^s_1(x) = c_2 z \Bigr) \Pi_{01}(x)= c_2 z
\sum_{n=0}^{\infty} \Bigl[ \sum_{k=0}^n p_{nk} x^k \alpha^{n-k}
\Bigr] \Pi^{(n)}_{01} (\alpha)= z M_1(\alpha) + M_0(\alpha) +
O_{01}(x),
\end{equation}
where the dimensionless function $O_{01}(x)$ is of the order $x$
at small $x$; otherwise it remains arbitrary. Again the first term
should be shifted to the third type of terms, while other two
terms are to be shifted to the first type of terms.

Completing this exact splitting/shifting procedure in the
expansion (A13), and restoring the explicit dependence on the
dimensional variable and parameters (A12), one can equivalently
present the initial expansion (A11) as follows:

\begin{equation}
d(q^2) = \Bigl( {\Delta^2 \over q^2} \Bigr) B_1(\lambda, \alpha,
\xi, g^2, a) + \Bigl( {\Delta^2 \over q^2} \Bigr)^2 B_2(\lambda,
\alpha, \xi, g^2, a) + \Bigl( {\Delta^2 \over q^2} \Bigr)^3
B_3(\lambda, \alpha, \xi, g^2, a) + ... + d_3(q^2; \Delta^2) + ...
\ ,
\end{equation}
since the coefficients of the above-used expansions depend, in
general, on the same set of parameters: $\lambda, \alpha, \xi,
g^2, a$, etc. The invariant function $d_3(q^2; \Delta^2)$ is
dimensionless, and it is free of the power-type IR singularities;
otherwise it remains arbitrary. In the formal PT $\Delta^2=0$
limit it survives, and is to be reduced to the sum of the first
type of terms in the expansion (A11). In other words, it is a sum
of $d^{MPT}(q^2)$ and $d^{PT}(q^2)$ up to third order, which have
been defined in section IV. The generalization to the next
iterations is almost obvious, and one finally obtains expansions
(4.9)-(4.13) for the full gluon propagator.

Concluding, let us underline that the splitting/shifting procedure
does not change the structure of the NL iteration solution at
small $q^2$. It only changes the coefficients at inverse powers of
$q^2$ in the corresponding expansion. In other words, it makes it
possible to rearrange the terms in the initial expansion (A11) in
order to get it in the final form (A19). Also, in the $q^2
\rightarrow 0$ limit, it is legitimate to suppress the subtracted
gluon self-energy in comparison with the mass gap term in the
initial Eq. (A1). Nevertheless, as a result of the
splitting/shifting procedure, which becomes almost trivial in this
case, one will obtain the same expansion (A19) with only different
residues, as just mentioned above. It is worth emphasizing that
residues remain completely arbitrary (undetermined) in any case.

\end{document}